Research article

# Design and implementation of multiprotocol framework for residential prosumer incorporation in flexibility markets

Miguel Gayo-Abeleira [a,b,*], F.J. Rodríguez [a], Carlos Santos [c], Ying Wu [d], Yanpeng Wu [d], Juan C. Vasquez [d], Josep M. Guerrero [d]

[a] *Department of Electronics, University of Alcala, 28805 Alcalá de Henares, Madrid, Spain*
[b] *Department of Artificial Intelligence, SMARKIA Energy SL, 24002 León, Spain*
[c] *Department of Signal Theory and Communications, University of Alcala, 28805 Alcalá de Henares, Madrid, Spain*
[d] *AAU Energy, Aalborg University, 9220 Aalborg East, Denmark*

A B S T R A C T

The growth of distributed renewable energy in the electrical grid presents challenges to its stability and quality. To address this at the local level, flexibility energy strategies emerge as an innovative technique. However, managing these strategies in residential areas becomes complex due to the unique characteristics of each prosumer. A major challenge lies in managing communication among diverse devices with different protocols. To address these issues, a comprehensive framework is designed and implemented to facilitate prosumers' integration in flexibility strategies, addressing communication at various levels. The effectiveness of the proposed framework is demonstrated through its implementation in a real smart home environment with diverse devices. The framework enables seamless integration and communication between IoT devices and IEC 61,850-compliant power devices. This research presents a novel approach to address the challenges of managing flexibility strategies in residential areas, providing a practical solution for prosumers to actively participate in optimizing energy consumption and enhancing the stability and quality of the electricity system amidst the growing integration of distributed renewable energy.

## 1. Introduction

The increase of distributed renewable energy resources in the electricity grid is essential to meet commitments to reduce greenhouse gas emissions and reduce dependence on oil and gas exporting countries. The use of distributed renewable energy resources has numerous advantages, including a reduction in transport losses, local redistribution of economic resources, reduction in the use of fossil fuels, etc. [1] but it also poses a significant challenge, which is the unpredictability of generation [2].

A possible solution to mitigate the unpredictability of renewable generation is the use of bulk generation with fast ramp up, such as thermal power plants or hydroelectric power plants [3], but the global trend is proving that this solution is not the most appropriate way. On the one hand, droughts are causing unavailability in hydroelectric power plants due to the absence of water for their operation and in thermal power plants due to water scarcity for cooling purposes [4]. Furthermore, global risks have led to price increases and even restrictions on the use of gas, which means that it cannot be used or is too expensive to be used in thermal power stations [5]. The global trend, driven by various governments [6,7,8], is towards a reduction in peak electricity demand due to technical, economic, and environmental benefits [9]. In this context, demand side flexibility (DSF) plays a crucial role in balancing consumption and generation in order to achieve grid quality in accordance with the most stringent standards.

* Corresponding author.
 *E-mail address:* miguel.gayo@uah.es (M. Gayo-Abeleira).

DSF can be defined differently based on the controllable resources within households. In the case where only loads can be controlled, the definition is equivalent to demand response (DR..), the adjustment of electricity use in response to existing generation through dynamic pricing or incentives. However, in its broader definition, it also includes prosumers, i.e., electricity users who not only consume energy but also generate and/or store energy [10].

The participation of prosumers in flexibility markets is particularly important in residential areas. In these locations, each prosumer has relatively small energy consumption or generation, but since there are numerous prosumers in the same area with similar energy profiles, grid-level effects are highly significant [11]. Users within the same geographic area can join together in renewable energy communities (RECs) and take advantage of their greater aggregate importance to the power grid. This grouping brings benefits to both the users and the grid operator, as these grouped users can assist in resolving voltage or frequency issues in the grid [12] generated typically due to the increase in renewable energy generation [13].

By implementing DSF techniques in residential areas, each prosumer is primarily looking for a reduction in his electricity bill [10]. If coordination between the different prosumers in the same area is not employed, the self-interests of prosumers can result in a failure to achieve the expected outcomes when implementing DSF techniques [14,15,16] e.g., shifting too many loads or generating too much energy to the most profitable hours. Therefore, with a view to achieve the expected benefits through the application of DSF techniques in residential areas, there must be coordination. This is achieved by prosumers sharing information among themselves and with the distribution system operator (DSO) or transmission system operator (TSO). Currently, these information flows necessitate complex multi-entity communications among the various stakeholders. This article seeks simplify the integration of all prosumer resources i.e., smart loads, distributed generation, and energy storage systems, in a transparent manner for users into local flexibility markets. Making it straightforward for prosumers in residential areas to join local flexibility markets would promote the adoption and integration of renewable energy generation [10].

With the aim of simplifying these communications, this work proposes a distributed hierarchical flexibility management strategy, where decisions regarding local prices and power consumption profiles are determined at the REC level. Each prosumer controls their respective devices through a home energy management system (HEMS). Coordination between the different prosumers is achieved through the creation of a blockchain (BC) network, which acts as the aggregator for the REC. BC is a decentralized information network that ensures the immutability of the data stored, ensuring that this data is accessible at every node of the network. Communications between prosumers and their resources are achieved through the FIWARE platform and the IEC 61,850 power grid communications standard. FIWARE is an open-source platform that enables the seamless integration of various IoT devices [17] while IEC 61,850 standardizes communications between electrical devices, including distributed energy resources (DERs) [18]. The whole system is based on the concept of modulation through virtualisation of different components. There are modules in charge of monitoring the household's resources, forecasting the load profile, managing the resident welfare, operating the resources or BC, among others. All these modules are integrated into a physical device in every household in the REC. This concept empowers the prosumer, allowing them to choose which functions to use in their household.

*1.1. State of the art*

Several research studies have been presented in recent years focusing on DSF at the neighbourhood level [10,19,20]. This is because if each household employs DSF techniques with its own benefit in mind, it may have counterproductive effects such as shifting the peaks to different times, but with the same problem as before applying the techniques [14,21].

Research by [22,23] reviewed methods for demand flexibility in residential buildings modeling and quantification. In [24] and [25], special attention is given to existing control and management approaches to exploit demand-side flexibility, but the multi-entity interactions were neglected. According to [26], there are two types of interaction between the different roles, coordination, and negotiation. In [27], an example of negotiation between different households and the utility is presented. Techniques for coordination between different households are described in [14] and [28].

In [29], a peer-to-peer energy market has been implemented between users and demonstrates that BC technology is valid for the operation of decentralized markets. According to [30], the BC is the emerging technology that is having the greatest influence on the application of flexibility techniques. It has been demonstrated that the application of flexibility techniques using BC results in energy and cost savings by avoiding manual operation on end devices [31]. One of its main problems for large-scale implementation of BC technology in energy markets is the massive amount of data it generates. This problem has been mitigated using aperiodic market techniques in [32]. According to [33] BC technology can be used to reduce peak demand. In this study, the results are validated through their implementation in 8 households in Canada, where peak reductions of up to 62% are achieved. In [34], BC technology is used to let the aggregator know the current status of each prosumer and thus, act accordingly by adapting the grid topology. In [31] a large number of research papers that have employed BC technology to implement DSF techniques are discussed. BC technology is used in the DSF literature, but these works do not address how communication with the end devices takes place.

The emergence of smart Internet of Things (IoT) devices has highly favoured the various activities targeting the DSF in the downstream residential sector, which is reforming towards utilizing massive data for operations and services. With the support of IoT, domestic appliances, various smart devices, sensors, and actuators are participating in the grid edge energy management systems, which can effectively increase local consumption, optimize scheduling of multi-energy resources and loads, improve living experience, and enable fast demand side response.

Huge amounts of consumption data are generated from various demand side applications, which actively engage customers to find the most cost-effective ways to support the transition to climate neutrality. Distributed energy assets can easily generate over one terabyte of data one day and coming from different communication protocols, it is necessary to have an aggregating and interactive IoT platform to achieve cross-domain interoperability. FIWARE, as one of the most popular open-source platforms generated in Europe from the Future Internet Public Private Partnership, is used to manage the integration of the large-scale and heterogeneous IoT systems with data privacy, economy, and reliability [17]. FIWARE supports many IoT communication protocols with IoT agents, such as IoT Agent for JSON (a bridge between HTTP/MQTT and NGSI), IoT Agent for LWM2M (a bridge between the Lightweight M2M protocol and NGSI), IoT Agent for UL (a bridge between HTTP/MQTT and NGSI with UL2.0 payload), IoT Agent for LoRaWAN (a bridge between the LoRaWAN and NGSI) [35]. With FIWARE, IoT ecosystems can create communities through peer-to-peer connections with common access to the smart energy components that they are operating via 'context broker' by processing real-time data in a highly decentralized and large-scale manner.

Nevertheless, power devices, which constitute a fundamental part of the equipment employed to manage flexibility tasks, typically lack compatibility with IoT communication protocols and instead adhere to specific electricity protocols, such as IEC 61,850. IEC 61,850 is a group of standards that was devised for the automation of electrical substations [36]. In recent years this standard has been extended to cover the needs of hydroelectric [37], wind power plants [38] and DERs [18]. Following the emergence of extensions to meet the needs of DERs, IEC 61,850 has attracted the attention of researchers, as it can provide standardized communication and control interfaces to achieve interoperability between all components of the electrical systems made up of DERs [39].

The IEC 61,850 set of standards is based on three pillars: a communication protocol, a generic object data model, and a description language for the configuration of Intelligent Electronic Devices (IEDs) and other information models [40]. Not only the codes and numerical values are transmitted, but the semantics of the respective data is also specified, something that is not found in prevalent protocols. The data is organized into tree structures that produce a tree data model [41]. The objects are defined using this generic model based on the project plan. Each unit is characterized as a logical device (LD). The LD is made up of appropriate logical nodes (LNs), that are ready-made groupings of data objects (DO) that fulfill specific functions. The data model together with the services of reporting or data setting, device control, file uploading or downloading and log transmission, among others, constitute the ASCI communication interface [42]. The ASCI is a self-contained block. The different IEDs that make up the system, and their interrelationship, are described using the SCL language [43]. In [44], the mapping between ASCI and Manufacturing Messaging Specification (MMS) is defined to allow data exchange over networks using the TCP/IP protocol. The standard also defines GOOSE messages, for fast transfer of event data for a peer-to-peer communication mode. IEDs can act as both servers and clients, so that peer to peer communication is achieved.

In [39], an EMS is designed and implemented based on the IEC 61,850 standard, resulting in demonstrating its effectiveness in both islanded and grid-connected microgrids. In [45], the feasibility of an IEC 61,850-based solution for electric vehicle charging and discharging management in photovoltaic (PV) installations is developed and demonstrated.

To the best of the authors' knowledge, the only proposal for the use of the IEC 61,850 standard in demand response scenarios or DSF is set out in [46], which combines IEC 61,850, OpenChargePoint protocol (OCPP), OpenADR and UDP which does not allow compatibility with whatever IoT device is in the household, and the results are only validated by simulation and not against real devices.

*1.2. Main contributions*

This paper proposes a framework to facilitate the engagement of residential prosumers in flexibility markets in an interoperable way, integrating IoT devices with IEC 61,850-based power equipment. The proposed framework enables prosumers to contribute their power absorption or injection capabilities using devices with IoT communications and devices compliant with the IEC 61,850 power grid communications standard. Moreover, it enables the establishment of local flexibility markets using BC technology to coordinate prosumers within the same geographic area and facilitate their collective participation in flexibility markets with DSO/TSO. Additionally, the framework empowers end users by enabling them to specify their preferences through a smartphone app.

For evaluating the performance in flexibility markets of the proposed framework, a realistic test platform has been developed, integrating many IoT devices, representing advanced smart home. In addition, it has also been possible to test the inclusion of IEC 61,850 devices by using a real time digital simulator (RTDS) to emulate a PV installation with storage capacity. Thanks to this test platform, the behavior of the IoT devices has been tested with different DSF strategies and different distributed generation systems. The test platform enables the emulation of a flexibility market scenario where the aggregator sends power references to prosumers based on the REC status.

By empowering prosumers to leverage a diverse array of devices for power absorption or injection, this research lays the foundation for enhanced coordination and active engagement of prosumers in flexibility markets, leading to increased efficiency.

In summary, the primary contributions of this study are as follows:

1. The authors propose a framework to facilitate the engagement of residential prosumers in flexibility markets in an interoperable way, integrating IoT devices with IEC 61,850-based power equipment.
2. The framework is designed to seamlessly integrate IoT devices and IEC 61,850 devices with BC technology, leveraging the potential of smart contracts and distributed ledgers.
3. The proposed concept offers a simple and user-friendly approach to integrating residential prosumers into flexibility markets, following a plug-and-play philosophy.

4 A test platform has been developed based on the framework, allowing the evaluation of distributed generation installations against real loads of a Smart Home.
5 The study includes the emulation of a flexibility market, where the aggregator provides the prosumer with a power reference based on the REC's status. The prosumer operates in real time on the available devices, validating the effectiveness of the design framework within an REC environment.

## 2. IoT and IEC 61,850 smart connected home concept

This work presents a scenario that closely resembles reality, considering the existing IoT devices in smart homes (such as smart appliances, smart plugs, sensors, etc.) that provide information on energy consumption and the current state of the household. Additionally, it considers IEC 61,850-compliant power devices responsible for generating, storing, and managing energy and their connection to the main grid.

Different roles are established, as depicted in Fig. 1: IEDs, prosumers, aggregators, and DSO/TSO. IEDs refer to devices with added communication capabilities that can be monitored and controlled. The prosumer role represents the end user interacting with the framework. The aggregator serves as the coordinating body for the REC. To enable the framework, a home energy management system (HEMS) is implemented on a single board computer (SBC), acting as the central communication hub among the different roles. The HEMS allows prosumers to input their preferences, including energy services, privacy, and health considerations, while also providing real-time information about the installation. Communication between the HEMS and the household devices, which can be IoT devices commonly found in smart homes or electrical energy generators and storage devices, enables the retrieval of updated information and the determination of each device's operational status. Finally, the HEMS establishes bidirectional communication with the aggregator to report power exchange between prosumers and the grid and to accommodate any necessary changes to their consumption patterns.

The DSO/TSO and the aggregator engage in communication to provide status updates on the REC and participate in collective-level flexibility markets. The aggregator determines the energy exchange between the REC and the main grid based on the information received from the DSO/TSO.

### 2.1. Modules

The HEMS comprises multiple virtualized modules that gather information from the prosumer via an interactive app, as well as from the aggregator and the physical devices within the prosumer's installation. These modules process the collected information based on the prosumer's preferences, instructions from the aggregator, and the current status of the devices. Subsequently, the modules transmit the device status and setpoints to establish new states that align with the requirements of both the prosumer and the aggregator.

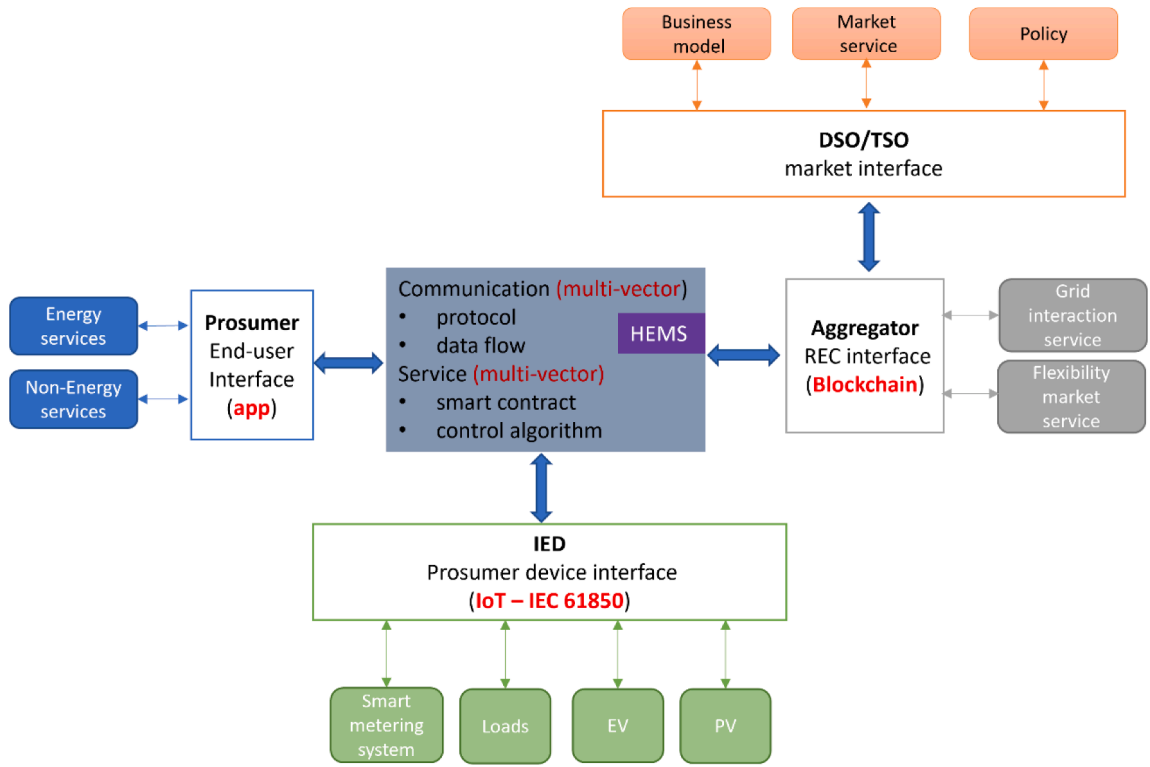

**Fig. 1.** Multi-vector communication architecture scheme, showcasing the HEMS as the central communications hub linking the prosumer, IEDs, and the aggregator. The aggregator facilitates communication between all HEMS systems of each prosumer and the DSO/TSO.

For instance, in the welfare management module, the user defines desired temperature settings for when they are at home or away. The DR.. management module utilizes this information, along with data from presence sensors in the advanced metering module, load expectations from the load profile module, and signals from the flexibility market via the BC module, to operate the temperature control devices.

*2.2. Prosumer role*

The primary role considered is that of the prosumer. The prosumer engages with the HEMS within their home to define their preferences. This interaction occurs via a web or mobile application. Within their preferences, the prosumer can specify various comfort parameters for the home (such as temperatures, humidity, schedules, etc.), device availability, and energy usage preferences (e.g., renewable energy, local energy, etc.).

The prosumer's consent is always required for the control of all household devices.

*2.3. Aggregator role*

Communication among prosumers is facilitated through the establishment of a BC network. In this context, the BC takes on the role of the aggregator. To accomplish this, each prosumer functions as a node within the BC network. This level manages the flexibility functions of each prosumer. These functions are implemented through the use of smart contracts (SCs). Various types of SCs can be developed, including those for local energy markets or flexibility markets. Several references in the literature demonstrate bidirectional communication between the aggregator and the prosumer utilizing BC technology [47,48,49].

Fig. 2 illustrates the flow of information between prosumers and the aggregator. The local flexibility market initiates by prosumers transmitting data to the BC through the execution of an SC. The prosumers who have signed the SC agree upon the quantity and level of detail of the transmitted data. The minimum required data from all prosumers is their anticipated power consumption profile. Upon receiving the power consumption profiles of all prosumers, the aggregator gains insight into the present state of the REC. Based on the REC status and the instructions from the DSO/TSO, another SC is executed, entailing the transmission of instructions to each prosumer.

*2.4. IED role*

Communication with the devices is the most complex to deal with in a realistic environment. In real-world scenarios, households possess diverse devices, each adhering to different communication standards. This paper's scope centres on the communication between prosumers and their devices, necessitating a more in-depth examination of these information exchanges.

Fig. 3 depicts the combined usage of the FIWARE platform and the IEC 61,850 communication standard. The FIWARE platform has been employed to consolidate data from IoT devices and smart home appliances. To address power devices, the IEC 61,850 standard has been selected as the most promising option for widespread adoption in Microgrids [50]complemented by the addition of an interface to FIWARE to integrate their information as well. FIG. 3 illustrates the communication schematic between the HEMS and the devices within the prosumer's installation, highlighting the message exchange between different components of the HEMS.

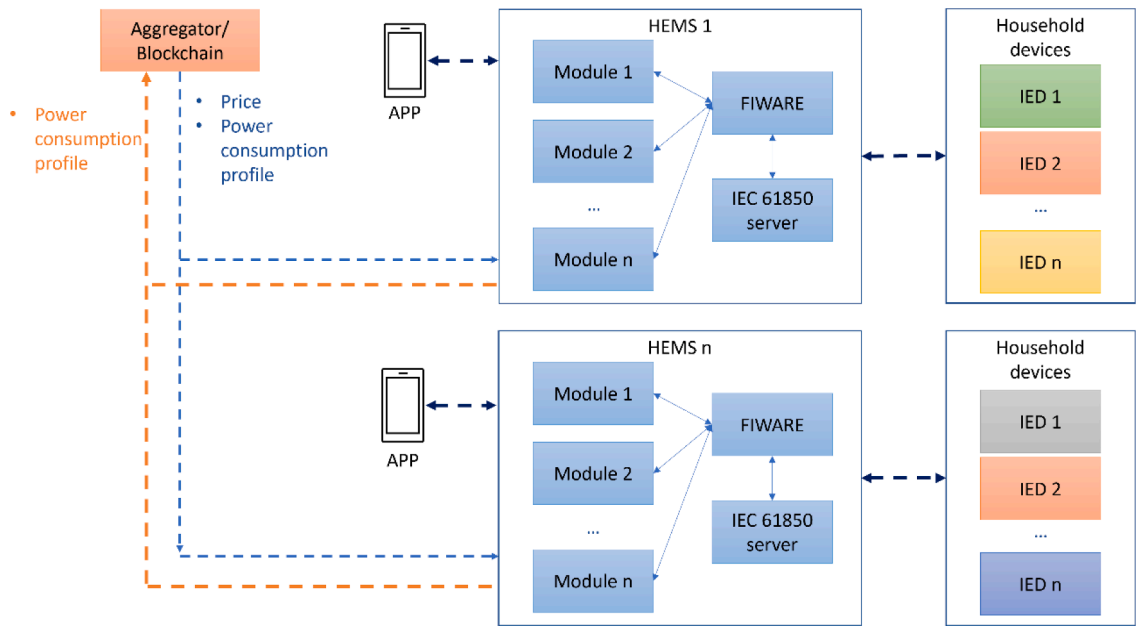

**Fig. 2.** Communication structure illustrating the connections between HEMS of prosumers and various roles.

The FIWARE framework is developed based on the IoT-A standard reference architecture, serving as a connector and organizer of IoT devices and enabling the rapid development of scalable applications [51]. This framework plays a crucial role in HEMS, as depicted in FIG. 3. FIWARE offers extensive compatibility with various IoT protocols and devices. Moreover, it provides interfaces for IEC 61, 850 MMS and REST services, facilitating seamless interaction between status data, control commands, the IEC 61,850 server, and other modules within HEMS.

Fig. 4 illustrates an implemented instance of an IoT architecture based on FIWARE for energy digitalization and automation, which is utilized in this particular case. As depicted in Fig. 4, IoT devices with diverse communication protocols are consolidated through gateways. Additionally, multiple generic enablers, such as IoT agents, are employed to collect valuable context information and leverage the robust FIWARE NGSIv2 API for communication with the Context Broker. The Context Broker serves as the central component within the FIWARE framework, facilitating the integration of various platform components to enable data processing, analysis, visualization, as well as data access control, publication, and monetization [35]. The IoT Agents function as protocol translation bridges, ensuring robust compatibility with diverse protocols and hardware devices.

Communications using the IEC 61,850 communications standard follow this process: Multiple IEDs exchange information with an IEC 61,850 server embedded in the physical household device using GOOSE messages, which store the status of all devices. For communication between the IEDs and the IEC 61,850 server, they need to be connected to the same local network through Ethernet cables. By publishing GOOSE messages on the local network, the IEDs notify the server about any changes in their status. The IEC 61,850 server disseminates GOOSE messages on the local network to instruct the IEDs to change their state. The IEC 61,850 server gathers data from all connected IEDs within the house and serves as the central point for exchanging information with the rest of the system. With the support of the IEC 61,850 MMS interface offered by FIWARE, bidirectional communication is established between FIWARE and the IEC 61,850 server. Consequently, FIWARE gains access to the data of all devices within the household and has the capability to send commands to modify their status accordingly. Once the IEC 61,850 server is integrated into the FIWARE framework, as depicted in FIG. 4, IEC 61,850 devices can seamlessly interact with FIWARE, similar to any other IoT device.

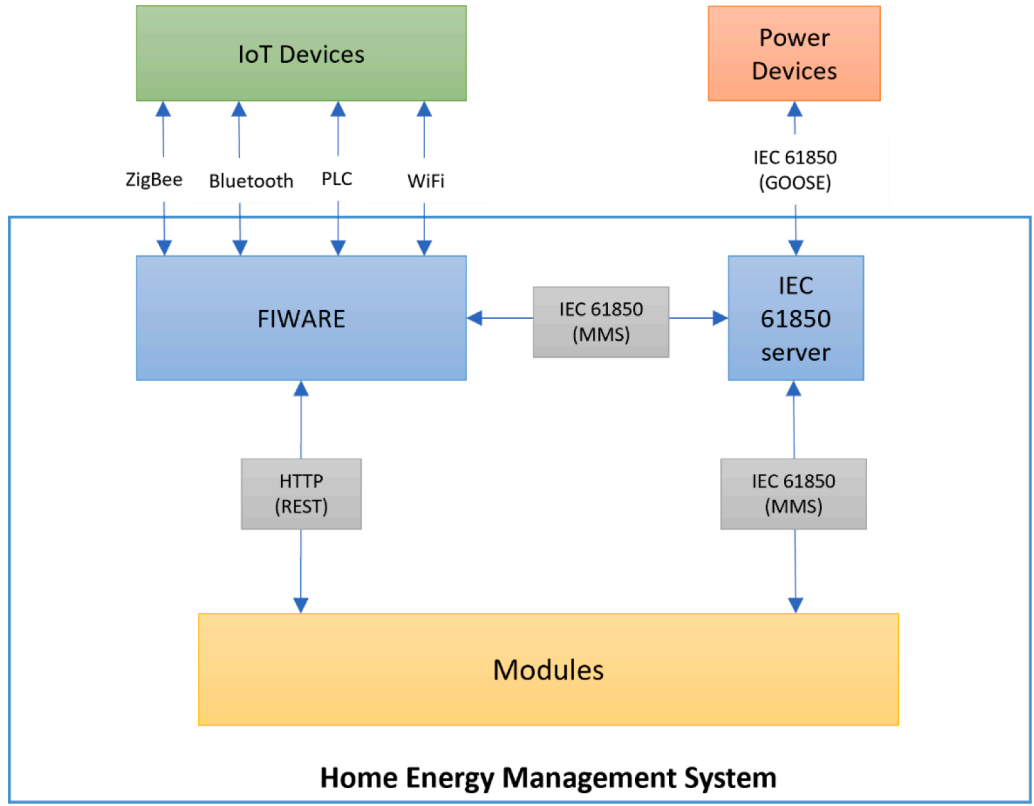

**Fig. 3.** Detailed representation of communication between HEMS and devices, as well as information transmission among HEMS components.

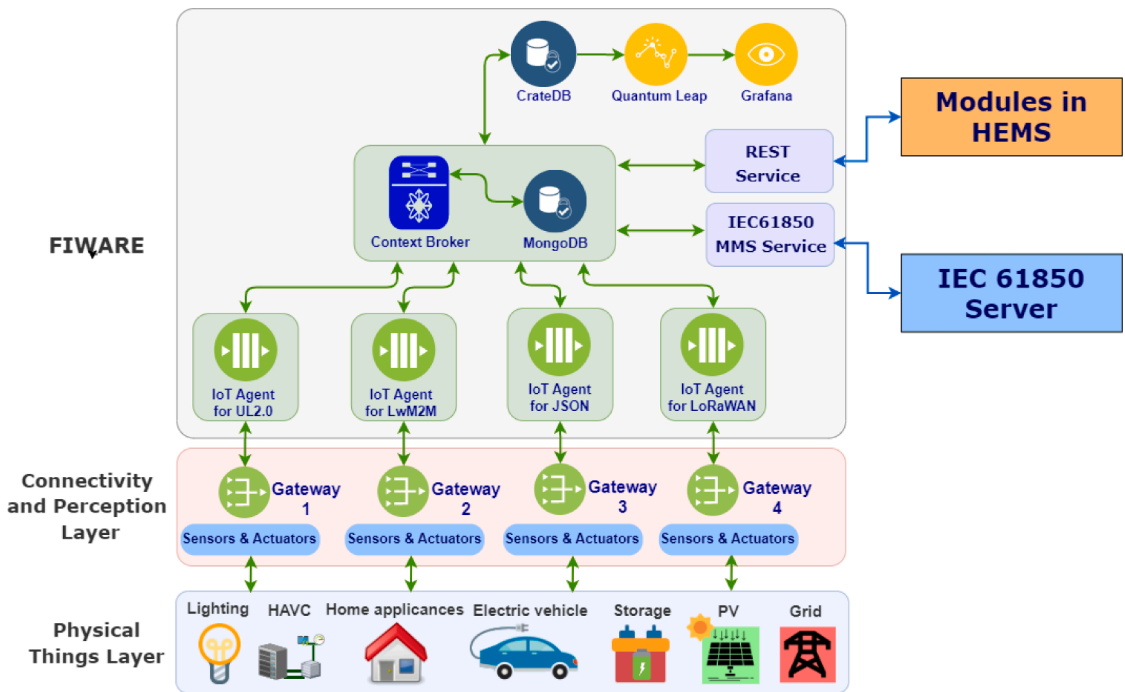

**Fig. 4.** FIWARE communication architecture for seamless integration of IoT devices and data management.

## 3. Test platform

In order to validate the effectiveness of the proposed framework, a test platform has been established utilizing Aalborg University's IoT lab [52] as a representation of a smart home. The IoT lab is equipped with a variety of appliances, sensors, and smart plugs featuring IoT technology, enabling comprehensive monitoring and control of device statuses. The RTDS is utilized to emulate a self-consumption PV installation with a battery, employing IEC 61,850-compliant devices. The test platform accurately replicates a prosumer's real home environment, incorporating domestic devices from various manufacturers and a self-consumption PV installation. The HEMS interacts with both the IoT lab and the self-consumption installation, collecting data and managing the devices accordingly.

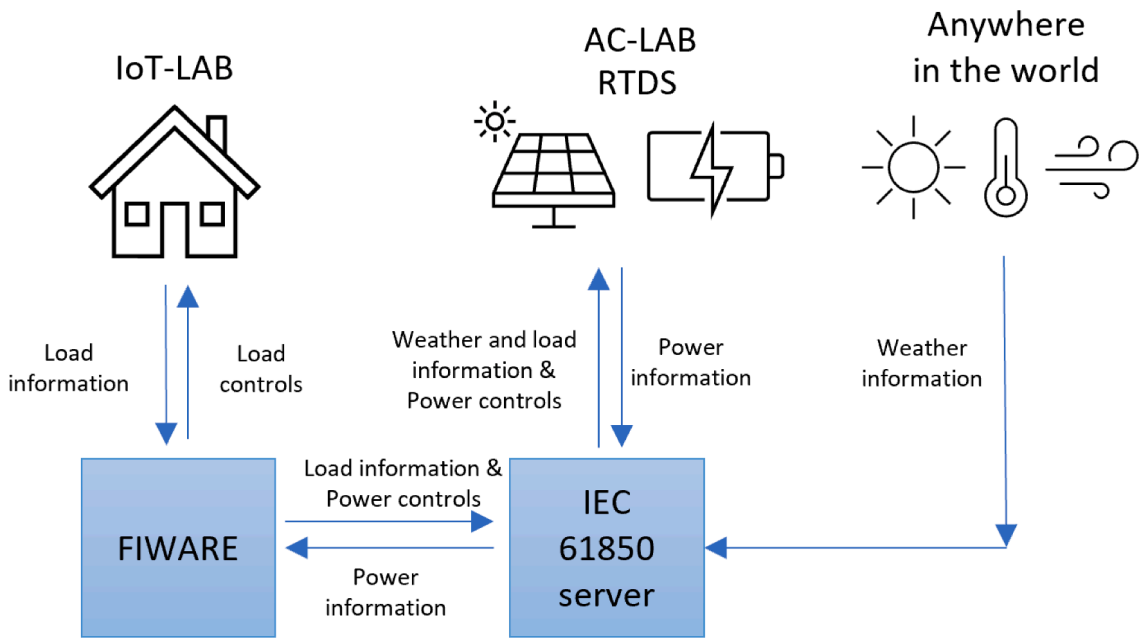

**Fig. 5.** Experimental framework for testing smart devices in diverse distributed generation installations under varying climatic conditions.

To ensure the realistic functionality of the PV system, accurate panel temperature and irradiance data are required. The RTDS is configured to enable real-time input of climatological data from internet-connected weather stations. This capability facilitates tests where the prosumer's virtual home can be located anywhere in the world, enabling assessments of DSF algorithms' performance under varying climate conditions. Fig. 5 depicts the architecture employed in the experimental framework.

### 3.1. IoT laboratory

The screen displayed in Fig. 7 illustrates the floor plan of the IoT-MG lab (Fig. 6), depicting the status of various devices. The toaster depicted in Fig. 7 serves as an example to demonstrate the implementation of load control using the proposed communications architecture. In this case, the toaster is operated through a smart plug, enabling remote activation and deactivation of the device.

### 3.2. IEC 61,850 server

According to [53], it is stated that HEMS can be fully modelled using standard IEC 61,850 LNs and DOs. Based on this premise, an IEC 61,850 server has been created to represent the entire installation for testing purposes. The server incorporates a data model of the installation using standard LNs and DOs. To implement this, the libiec61850 library [54], written in C language, has been utilized. The IEC 61,850 server has been deployed on an SBC, specifically a Raspberry Pi 4 Model B [55].

The IEC 61,850 server acquires real-time load data from FIWARE, weather data from the Thingspeak[56] cloud database, and the reference battery power data is calculated by the Raspberry Pi, functioning as a battery management system, based on the aforementioned data.

Regarding the load modeling from the perspective of the IEC 61,850 server, the predefined LN MMXU has been employed. This LN enables the monitoring of alternating current power measurements associated with load energy consumption within the installation. The battery has been modelled using LNs ZBAT and ZBTC. The former stores battery system characteristics, actual operating parameter measurements, and facilitates the transmission of control commands to switch the battery on and off. The latter contains the necessary data for monitoring and controlling the battery charger. For the PV generation system, the following predefined LNs have been utilized: MMDC, MMET, MMXU, and ZINV. MMDC encompasses DC power measurement information, MMET stores weather condition data, MMXU provides power measurement and operational status of the inverter, and ZINV allows configuration of various parameters related to the inverter.

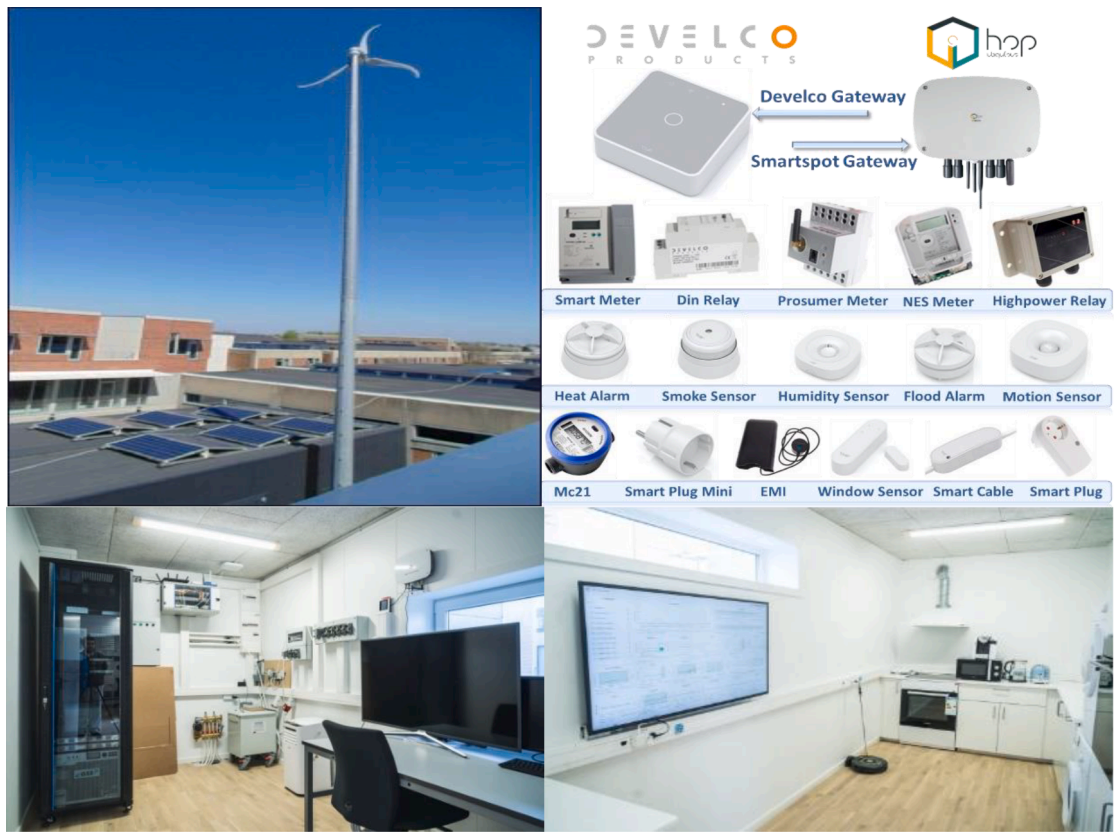

**Fig. 6.** IoT-MG lab showcasing IoT-enabled loads, RESs, sensors, and smart plugs.

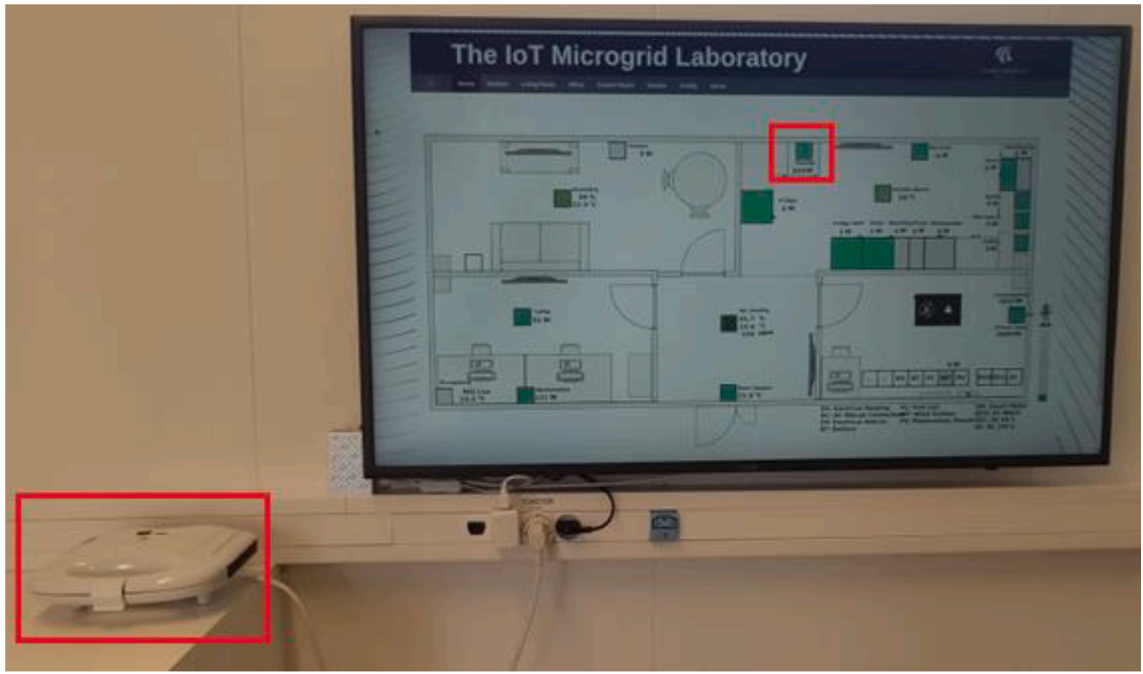

**Fig. 7.** Controllable 700 W load via smart plug located in the IoT-MG lab.

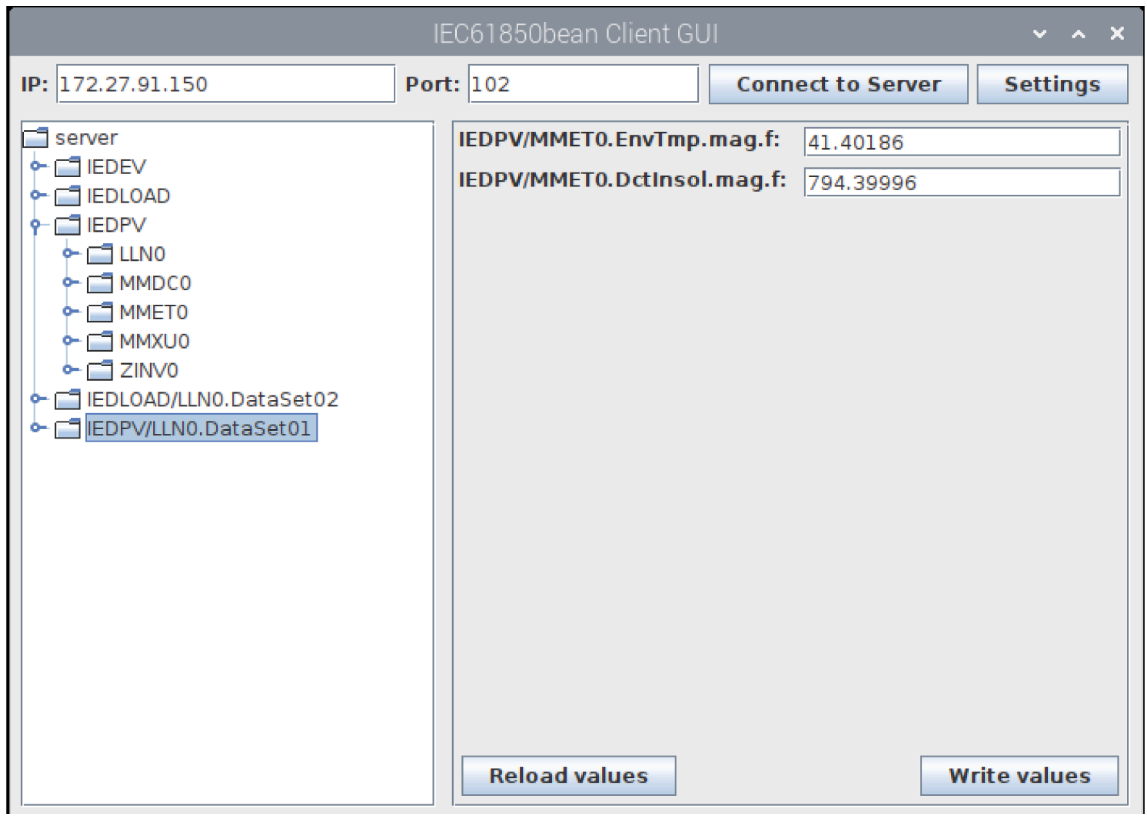

**Fig. 8.** Structure of the data in the IEC 61,850 server employed in the experimental framework.

FIG. 8 depicts a screenshot of an IEC 61,850 client that is connected to the previously configured server. The left side of the image displays the data structure, providing an overview of the information organization. On the right side, the client displays the specific details of the weather data it is retrieving from Thingspeak, showcasing the real-time weather information being accessed.

### 3.3. RTDS configuration

In RTDS, the emulation of a PV installation with a battery and a load has been implemented using the RSCAD software. This emulation setup is connected to the main grid and requires various data inputs for proper operation. Firstly, it requires the active and reactive power values of the load. Secondly, it relies on weather data to simulate the PV panel generation. Lastly, it needs the battery power reference to accurately model its behavior. To facilitate the provision of real-time operational data, the emulated installation has been configured as an IEC 61,850 server. This emulated server subscribes to the data from an externally created IEC 61,850 server that contains all the necessary data. As a result, the real server broadcasts GOOSE messages automatically whenever there is a change in the data, ensuring the emulation operates correctly. GOOSE messages are directly sent through the Ethernet physical layer, enabling latency times of less than 3 ms to be achieved.

Fig. 9 illustrates the experimental setup. The red box labelled 1 represents the RTDS system, while the box labelled 2 corresponds to the RSCAD program utilized for designing the emulation, transferring it to the RTDS, and monitoring real-time operation. The Raspberry Pi 4, indicated by label 4, is connected to a screen labelled 3, which displays the messages generated by the IEC 61,850 server. Lastly, the switch labelled 5 is employed to integrate the RTDS, the workstation with RSCAD, and the Raspberry Pi 4 into the local network.

Even though a specific installation has been emulated, it is possible to modify the emulation in RTDS to create a different installation according to specific requirements. To adapt the PV installation with a battery (shown in Fig. 10) to another configuration, one can adjust parameters such as the number and type of panels, as well as the capacity, charging, and discharging power of the battery. By making these changes, the IoT lab can simulate a residential energy consumer (REC) located anywhere in the world with a customized electrical setup. This flexibility enables the testing of DSF algorithms with various distributed generation installations and diverse climatic conditions, using real loads as a basis.

For this specific case, the emulated electrical system matches the one already installed at the Polytechnic School of the University of Alcalá, ensuring the realism of the emulation. In the actual installation, the irradiance and temperature parameters of a PV panel are monitored (as shown in Fig. 11) and uploaded to a Thingspeak cloud database. The IEC 61,850 server reads the same weather data and transmits it in real-time to the RTDS through GOOSE messages. The electrical system consists of PV panels with a peak power of 4 kW under ideal conditions (1000 W/m2 and 25 °C), and a battery with a maximum charging and discharging power of 1.8 kW and a capacity of 8 kWh.

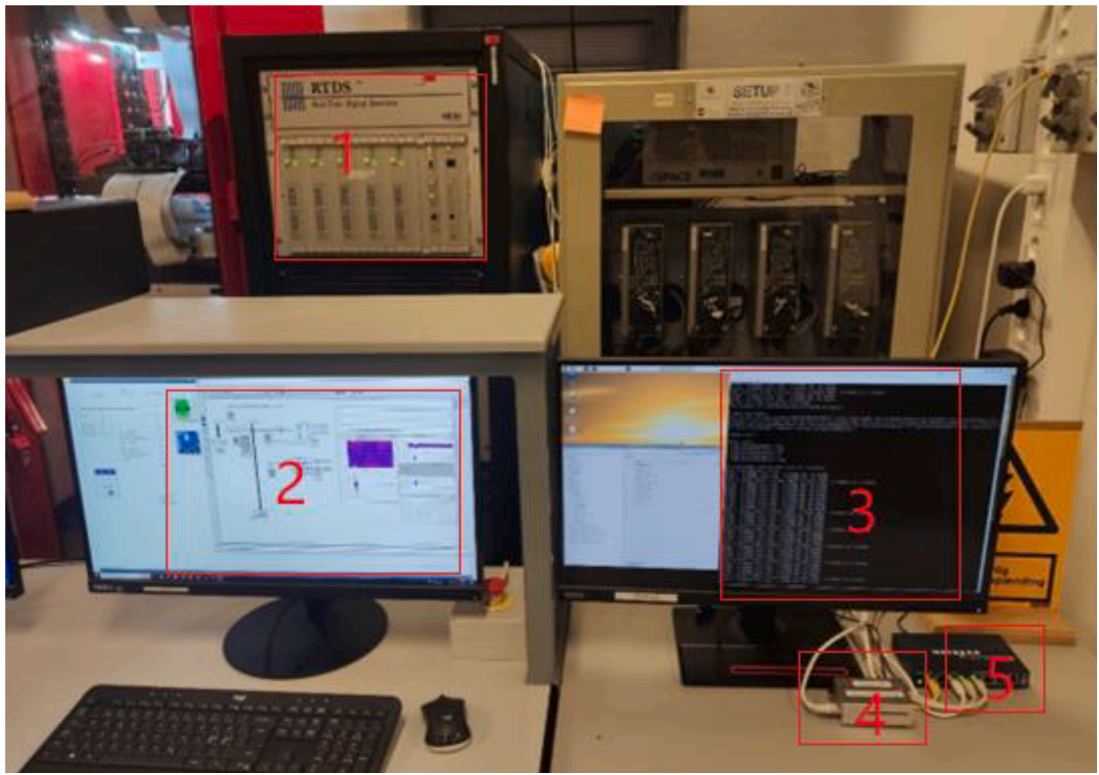

**Fig. 9.** Experimental setup with RTDS as an emulator and Raspberry PI 4 as a virtual IEC server.

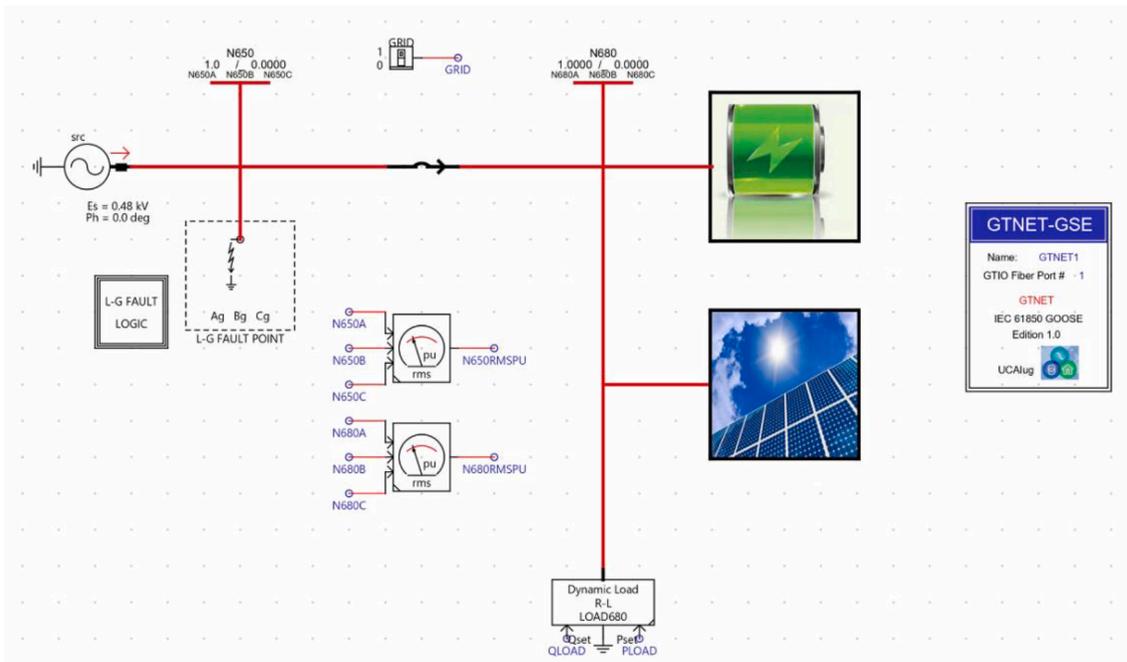

**Fig. 10.** Screenshot of the emulation in RTDS. Battery and PV generation are emulated using actual weather data, while all other data is acquired from the real installation.

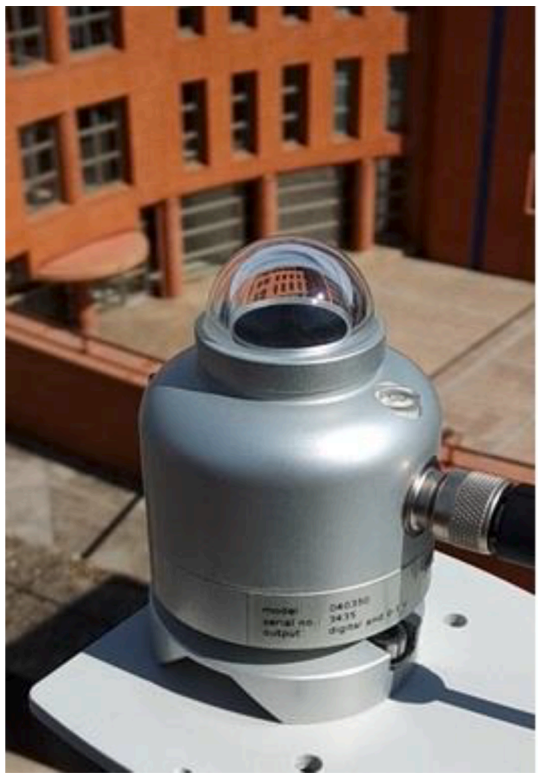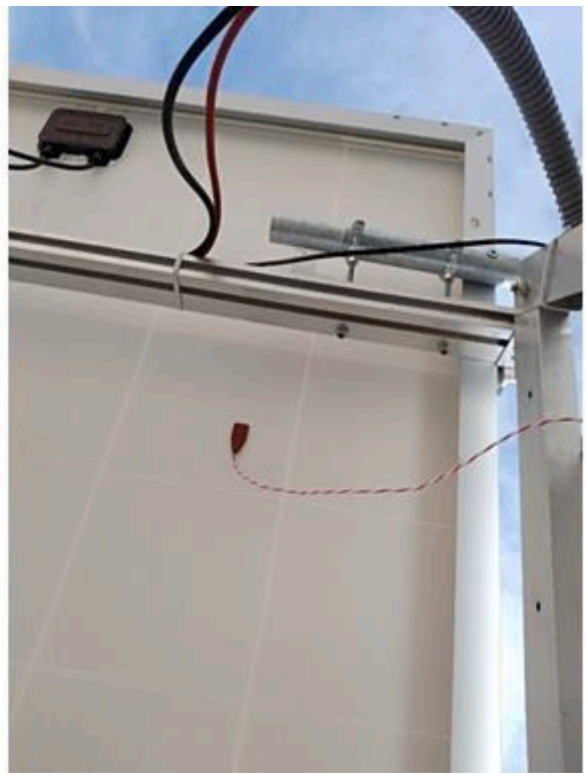

**Fig. 11.** Irradiance sensor (left) and panel temperature sensor (right) used to transmit weather data to the emulation in the RTDS in the tests performed.

## 4. Experimental results

This section presents the results obtained from the test platform described in Section 3. Initially, a simple example of a flexibility market condition is considered to demonstrate the effectiveness of the proposed solution on each device individually. The objective in this scenario is to minimize the energy fed into the grid by following a specific command. This scenario is particularly interesting because the test days have high irradiance levels with minimal cloud cover, resulting in an oversized PV installation compared to the loads. This leads to excess generation that needs to be absorbed by the grid. Since neighboring households in the same area typically have similar generation and consumption profiles, grid stability issues can arise. These tests were conducted over three different days, with each element being tested separately.

Once the effectiveness of the proposal has been confirmed, a more realistic flexibility market is defined, considering real-time power setpoints provided by the aggregator based on the current state of the REC. This allows for simulating coordination among the different prosumers within the REC.

The meteorological data used for the PV generation emulation, as explained in the previous section, is obtained in real time from the weather station at the University of Alcalá, which experienced clear skies during the test days.

Fig. 12 and Fig. 13 display screenshots of the GOOSE messages emitted from the Raspberry Pi 4 Model B, which serves as the HEMS, and from the RTDS, respectively. These screenshots capture the GOOSE messages using the Wireshark packet analyser.

**Fig. 12.** Screenshot of GOOSE messages emitted by HEMS in the test framework.

Fig. 13. Screenshot of GOOSE messages emitted by RTDS in the test framework.

These two screenshots display the exchange of information between the two devices through the utilization of GOOSE messages, which are specified in the IEC 61,850 communications standard.

### 4.1. Scenario 1: battery management

In the first test scenario, the focus is on a house equipped with a battery that has sufficient capacity to accommodate any surplus energy generated by the installation. By continuously monitoring the state of charge (SOC) of the battery, as well as the energy generation and consumption within the installation, a power reference is transmitted to the battery. The battery functions to supply energy when the demand exceeds the generation capacity, and it absorbs energy when the opposite occurs. As a result, the main grid is primarily utilized to handle abrupt power variations, while the battery rapidly responds by absorbing excess energy or fulfilling any energy deficits within the dwelling. This approach allows the PV system to operate at its maximum power point, while providing the user with the ability to consume the necessary energy to maintain their desired comfort levels. Essentially, the HEMS assumes the role of a battery management system, consistently providing instructions to guide the battery's actions.

In the Fig. 14, the graphical representation of the results obtained from the execution of scenario 1 is depicted. It illustrates the power profile of the loads in the IoT lab, with a baseline consumption of around 1 kW and occasional peaks of up to approximately 4 kW during short periods, corresponding to the activation of various household appliances.

From the start of the test around 8:45 a.m. until 10:00 a.m., the power consumption in the IoT lab exceeds the PV generation. In this phase, the battery supplies the additional power required to avoid drawing energy from the main grid.

Starting from 10:00 a.m. until the end of the test at 16:00, the PV generation surpasses the consumption, resulting in surplus energy. To prevent feeding excessive energy into the grid, as directed by the aggregator, the battery begins absorbing the excess energy. Throughout this period, there are intermittent instances of power exchange with the main grid. Two main reasons contribute to these exchanges:

1. Sudden power imbalances occur due to load variations or the presence of clouds, and the battery's capacity is insufficient to absorb such abrupt changes.
2. The battery reaches its maximum power capacity and can no longer absorb additional energy. During these critical moments, power exchange with the main grid becomes necessary to maintain the overall electricity quality within the dwelling.

During the test conducted in scenario 1, the total error with respect to the setpoint was less than 0.5 kWh. This small deviation can be attributed to the fact that the battery had sufficient capacity throughout the test duration. However, there were moments when the power demanded exceeded the battery's capability to supply. As a result, the actual power deviated slightly from the setpoint, leading to the observed error.

*4.2. Scenario 2: inverter management*

In the second test scenario, where there is no battery available in the installation, the power consumption and PV generation are monitored. If the PV generation exceeds the consumption, a signal is sent to the inverter to adjust its operation and generate power equal to the load's demand. The objective is to minimize power injection into the grid and rely on the grid only for sudden changes in consumption or generation.

The results of test 2, depicted in Fig. 15, show similar baseline loads as the previous day's test, around 1 kW. However, there is an increase in consumption between 10:30 and 12:15. During the first hour of the test and the period of increased consumption, the PV energy generated is insufficient to meet the demand. From 12:15, the PV generation becomes sufficient to cover the load until approximately 14:00. The generation adapts to the demand, resulting in minimal power exchange with the grid. However, the presence of clouds influences the test results, causing moments of sufficient generation and moments of insufficient generation. As a result, there are instances where power needs to be drawn from the grid to meet the load demand.

In the second test, the quantified error is measured as the amount of energy fed into the grid, which was calculated to be 0.269 kWh. As shown in Fig. 15, the deviation from the setpoint is mainly due to the periods when there has not been enough PV generation to meet the load, but in this test the aim is to inject as little as possible.

Fig. 16 provides a closer look at the results of this test, with a zoomed-in time scale. The purpose is to examine the details and understand why there are moments when energy is absorbed from the grid despite having enough irradiance to generate the required power. It can be observed that there is a load of approximately 400 W activated for a duration of one minute. Even though there is sufficient PV power generation capacity during that period, the inverter cannot instantaneously adjust to follow the load curve. As a result, for a brief period, power is absorbed from the grid. Similarly, when disconnecting the 400 W load, there is a delay in lowering the generation power, leading to a short period of power injection into the grid.

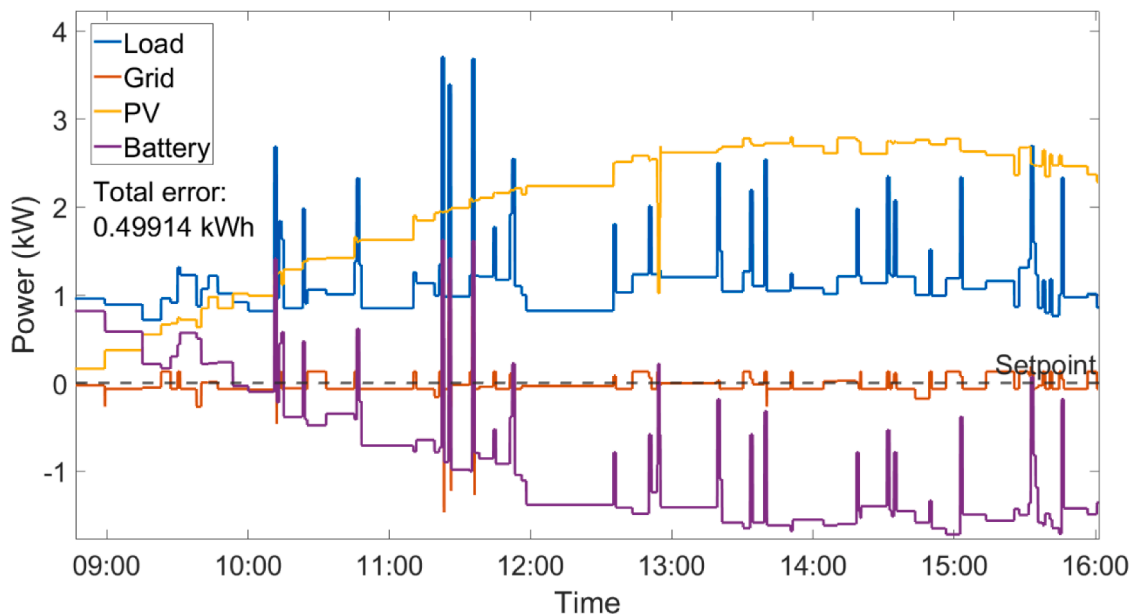

**Fig. 14.** Test results of near-zero power exchange achieved by controlling the battery power setpoint.

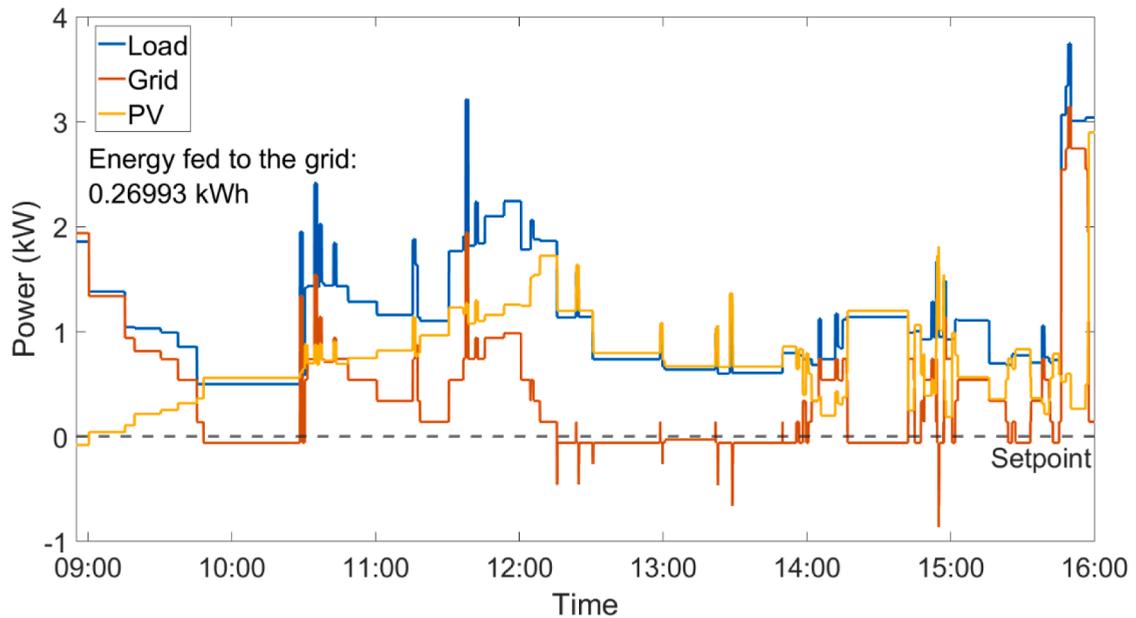

**Fig. 15.** Test results of near-zero power exchange achieved by controlling the inverter to deviate from MPPT and generate only the required power.

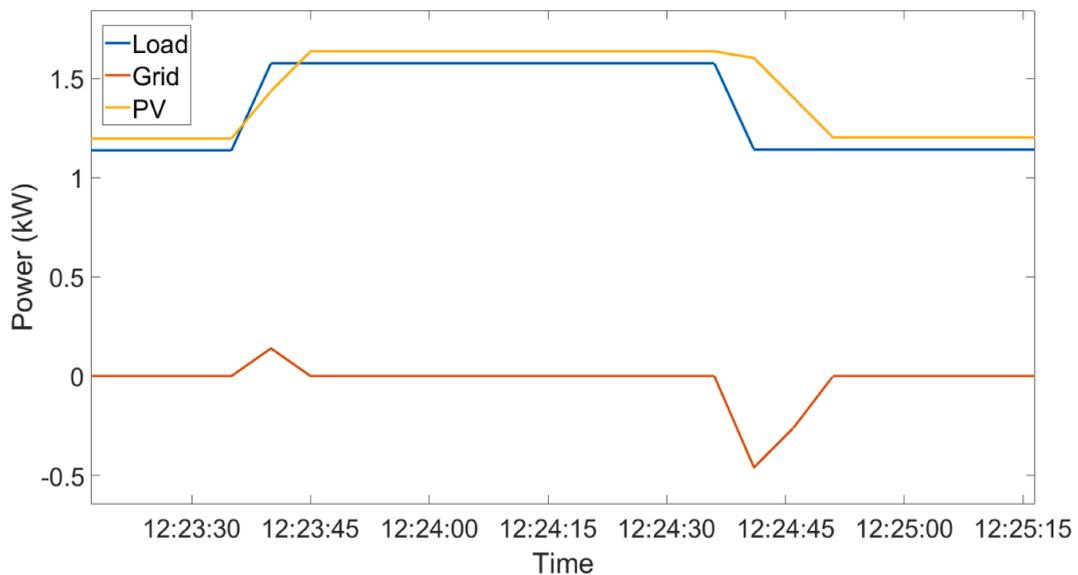

**Fig. 16.** Detail of the reaction times observed during the tests.

These short-lived deviations in power exchange contribute to the quantified error, as energy is injected into or absorbed from the grid for brief intervals.

### 4.3. Scenario 3: load management

In the third test, a self-consumption installation is considered, where no battery is present, and the inverter cannot deviate from the maximum power point tracking (MPPT) to control the generated power. Therefore, the only option to reduce surplus power injection into the grid is to activate controllable loads. In this specific test, a 700 W load controlled by a smart plug, represented by a toaster, is turned on and off. In a more realistic scenario, a water heater or a programmable appliance could be used to consume excess energy when available, avoiding grid consumption during periods of insufficient PV generation.

During the third test (Fig. 17), the load profile resembled that of the second test, with a baseline consumption of around 1 kW and an increase in consumption between 10:45 and 12:30. There was excess PV generation between 09:45 and 10:45 and from 12:45 to 16:00. During the first period after 10:30 and throughout the second period, the excess generation was significant enough to activate the 700 W controllable load, resulting in reduced power injection into the grid. This strategy helps utilize the excess energy by diverting it to the controllable load instead of exporting it to the grid.

In the third test, the total error with respect to the grid zero exchange setpoint was 6.66 kWh, which is the highest among the three tests. This was expected because the activation and deactivation of a single load does not provide precise control over the power exchanged with the grid. However, it still helped in reducing the error compared to not having any load activation, which would have resulted in an error of 8.98 kWh.

The results of the tests conducted over the three different days demonstrate the effectiveness of acting on various components of the installation to minimize energy injection into the grid during periods of excess generation. By effectively controlling the battery, the PV generation system, and the controllable loads, the designed architecture successfully achieved the desired outcome.

## 4.4. Flexibility market test

FIG. 18 shows the power profile of the REC (Residential Energy Community) with the main grid, representing the power exchanges between the REC and the grid. Positive values indicate power injections into the grid, while negative values indicate power absorption from the grid. The power profile is based on real data and represents the power exchanges of a residential area with 35 single-family houses, where 60% of the houses have a PV system installed, on a sunny day [34].

In the coordinated test, the REC aggregator provides a power reference every 15 min, taking into account the power exchanged with the main grid. This power reference is calculated considering the capacity reported by each prosumer (residential energy consumer) within the REC. The aggregator sends this power reference information to the HEMS (Home Energy Management System), which coordinates the operation of the PV system, battery, and controllable loads in response to the provided reference.

This coordinated interaction among the devices aims to optimize the power exchanges with the grid based on the real-time conditions and power requirements of the REC. By adjusting the operation of the PV system, battery, and controllable loads according to the power reference, the REC can effectively manage its energy consumption and generation, minimizing reliance on the grid and maximizing self-consumption.

In the coordinated test, the HEMS has communicated to the aggregator its exchange capacity availability with the grid, which is $\pm 2$ kW (positive or negative). To achieve this, the HEMS employs several strategies to regulate power exchanges and ensure that the exchange capacity remains within the desired range.

The HEMS activates and deactivates a controllable load with a power of 700 W for a total of two hours per day. This load helps to balance the power exchanges with the grid by consuming or supplying power as needed. By controlling the activation and deactivation of this load, the HEMS can adjust the power injection or absorption from the grid.

Furthermore, the HEMS regulates the operation of the battery and PV system. If there is excess PV power generation beyond what is necessary, the HEMS can instruct the inverter to reduce the generation power, thereby avoiding injecting unnecessary power into the grid. On the other hand, when the demand exceeds PV generation, the HEMS can utilize the battery to supply the additional power, reducing the need to absorb power from the grid.

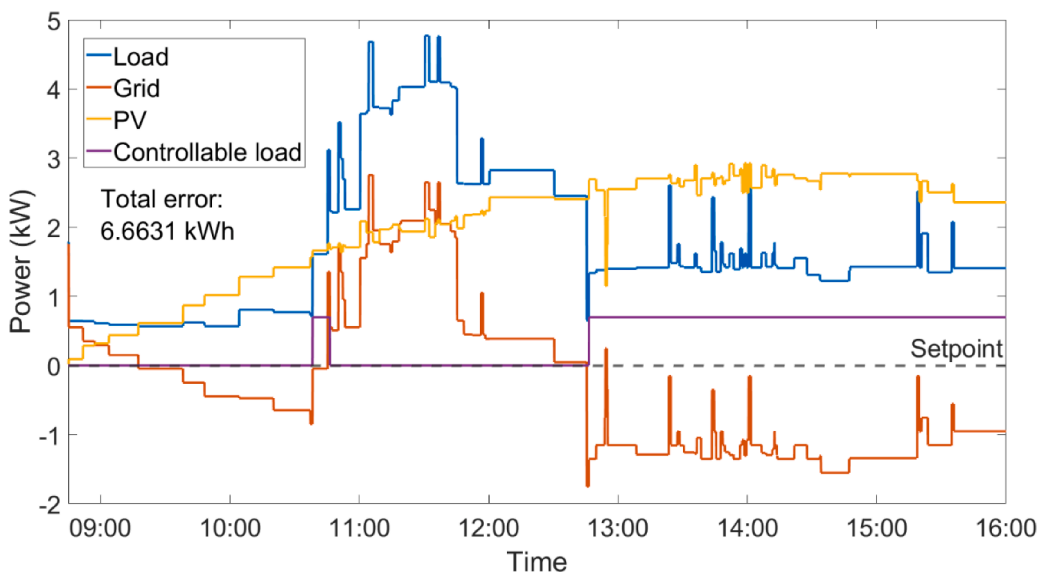

**Fig. 17.** Test result of near-zero power exchange test by switching a controllable load on and off.

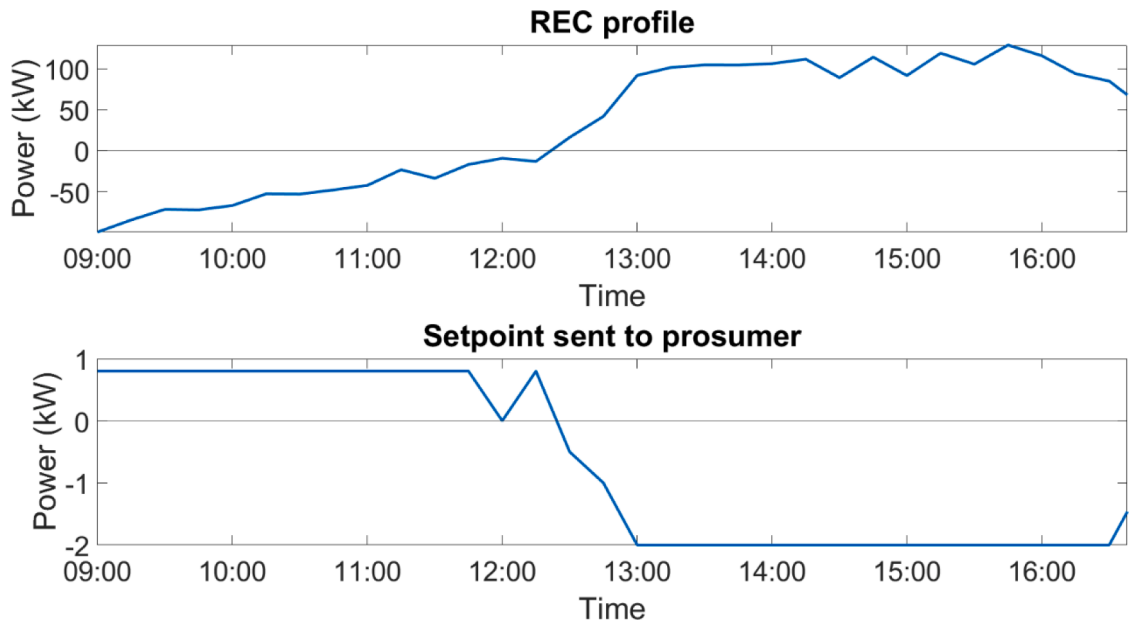

**Fig. 18.** At the top, power exchange profile of a residential REC of single-family houses and 60% of houses with solar installation on a sunny day [34]. At the bottom, profile of the setpoints that are sent to the prosumer based on the needs of the REC.

The profile of setpoints sent by the aggregator to the HEMS is shown at the bottom of FIG. 18. Positive power values indicate grid injections, while negative power values represent power consumption from the grid. These setpoints reflect the desired power exchanges based on the REC's overall power profile and the exchange capacity availability communicated by the HEMS.

By coordinating the operation of the controllable load, battery, and PV system based on the setpoints provided by the aggregator, the HEMS ensures that the REC's power exchanges with the grid align with the desired exchange capacity, optimizing self-consumption and reducing reliance on the grid.

In Fig. 19, the results of the coordinated test are presented, illustrating the power exchanges between the prosumer and the grid. Several observations can be made from the figure:

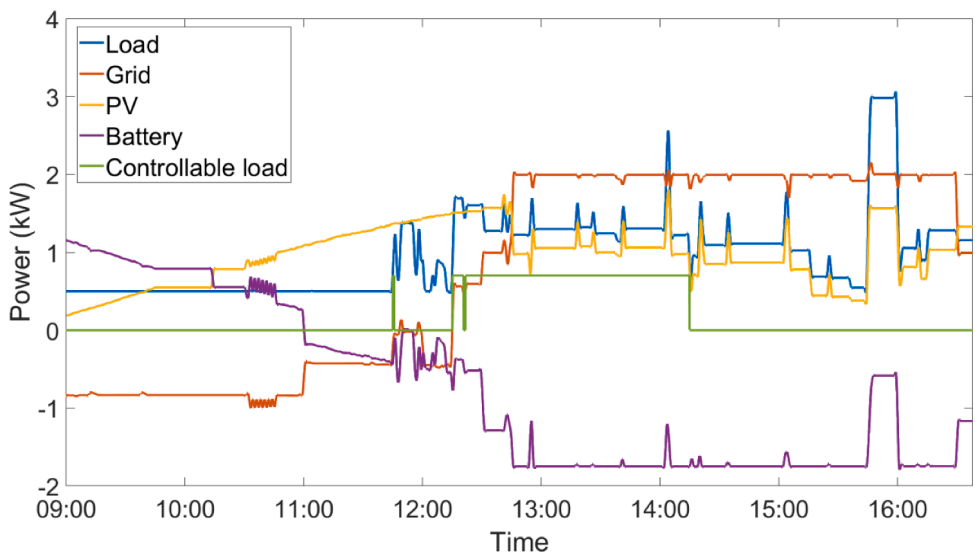

**Fig. 19.** Results of the test of power exchanges with the grid established by the aggregator.

1. During the initial hours of the test, the PV generation is insufficient to meet the power requirement for grid injection. Therefore, the battery is utilized to supply the additional power needed to maintain the desired power exchanges.
2. As the hours progress, the PV generation increases, reducing the dependency on the battery. At around 11:00, the battery starts absorbing power from the grid as there is surplus PV generation that exceeds the demand.
3. At 12:15, the controllable load is switched on and remains active for two hours. During this period, in addition to the battery absorbing power, PV generation must be reduced to achieve the target of absorbing 2 kW from the grid. This coordinated action helps to balance the power exchanges and optimize self-consumption within the REC.
4. The battery continues to absorb the maximum possible power from the grid throughout the rest of the test, except during periods when the house consumption increases. During these moments, the generated PV power is utilized, reducing the need for power absorption from the grid.

By coordinating the operation of the battery, PV system, and controllable load, the prosumer successfully adjusts its power exchanges with the grid based on the setpoints provided by the aggregator. This coordinated approach enables the prosumer to optimize self-consumption, reduce reliance on the grid, and actively participate in the flexibility market.

In the Fig. 20, the setpoint sent by the aggregator to the prosumer, the actual power exchanged with the grid by the prosumer, and the error between the setpoint and actual exchange are displayed. The following observations can be made:

1. The setpoints provided by the aggregator to the prosumer indicate the desired power exchanges with the grid, with positive values representing grid injections and negative values representing grid absorptions.
2. The prosumer's actual power exchanges closely follow the setpoints provided by the aggregator. There is a small deviation between the setpoints and the actual exchanges, which can be attributed to the response time and control dynamics of the devices in the prosumer's setup.
3. The error, which is the difference between the setpoint and actual exchange, is generally small throughout the test. Despite being requested to absorb energy when there is excess generation, the prosumer manages to follow the aggregator's instructions quite satisfactorily.
4. The total accumulated error over the duration of the test is 0.97 kWh. This indicates that the prosumer's power exchanges align well with the setpoints provided by the aggregator, with a relatively small deviation.

Overall, the results demonstrate the effectiveness of the coordinated approach in managing power exchanges between the prosumer and the grid. The prosumer successfully responds to the setpoints sent by the aggregator, contributing to grid stability, and optimizing self-consumption within the REC.

*4.5. Discussion of results*

Indeed, the results of the tests have demonstrated the feasibility and effectiveness of the designed framework, opening up possibilities for further enhancements and functionalities. The direct communication with power elements through the IEC 61,850 standard provides a powerful means to optimize power flows and achieve grid stability.

One interesting aspect is the utilization of PV panels as reactive power generating elements. By controlling the reactive power output of the PV installation, the prosumer can actively contribute to maintaining optimal grid conditions, such as frequency and voltage margins. This capability allows the prosumer to follow the instructions provided by the aggregator and participate in ancillary service markets, offering reactive power injection or modifying active power flows to support the DSO/TSO in maintaining grid quality levels.

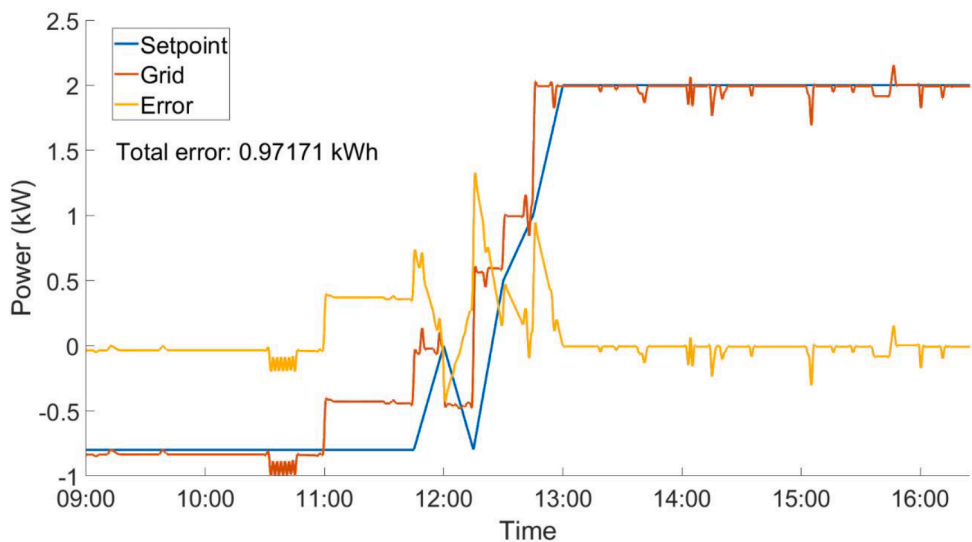

**Fig. 20.** Graph depicting the setpoint sent to the prosumer, the power exchanged with the grid, and the difference between these two parameters.

By enabling prosumers to engage in these flexibility services, the proposed framework simplifies their access to RECs and encourages their active participation. This not only benefits the prosumers themselves but also contributes to grid stability and the integration of renewable energy sources into the main grid.

The framework's ability to provide direct control and communication with power elements enhances the intelligence and adaptability of the HEMS. As a result, it enables prosumers to contribute to the optimization of power flows, improve self-consumption, and potentially participate in various grid services, unlocking new opportunities for the integration of renewable energy and the development of more sustainable energy systems.

The results obtained in this study have the potential to be applicable in any flexibility market, thereby offering the possibility of emulating weather conditions from any region worldwide. The main adaptations to consider would involve characterizing the specific communication requirements with the REC aggregator and complying with the legislative requirements for communication with the DSO/TSO in each particular market. As a result, the proposed framework provides a solid foundation for future research and development, enabling more efficient participation of prosumers in flexibility markets worldwide.

## 5. Conclusions

This paper proposes a multiprotocol framework that integrates IoT devices in a Smart Home with power elements compliant with the IEC 61,850 standard. The framework enables prosumers to participate in flexibility markets using their household devices in a user-friendly manner. The authors have created a test platform using the IoT lab at Aalborg University, which simulates a smart home environment with various smart appliances, smart plugs, and sensors. The platform also includes an RTDS for emulating power generation facilities and real-time communication.

Through the test platform, the proposed framework has been successfully tested, facilitating bidirectional communication between the HEMS and both the IoT lab devices and the power devices emulated in the RTDS. Three tests were conducted to achieve near-zero power exchange by controlling the battery, inverter, and a controllable load. Power exchange commands were also sent to the prosumer based on the status of a residential REC, demonstrating the coordination between prosumers in the REC.

In summary, this work introduces a comprehensive framework that seamlessly integrates IoT devices and power elements, facilitating efficient coordination and active involvement of prosumers in flexibility markets. Additionally, it highlights promising opportunities for future research and system improvement.

## 6. Future research directions

Future work includes several developments that will enhance the capabilities of the proposed framework and make it more accessible and appealing to end-users. One key area of focus is the development of a user-friendly application for end-users, which will provide an intuitive interface for interacting with the system and specifying preferences. This will make the system more accessible to a wider audience and increase its appeal to potential users.

In addition to improving the user experience, future work will also focus on enhancing the technical capabilities of the system. New modules for generation and consumption prediction will be added, providing more accurate and reliable forecasts of energy production and usage. This will enable prosumers to make more informed decisions about their participation in flexibility markets and optimize their energy consumption.

Another important area of future research is the integration of communication between prosumers and the REC aggregator. This will allow for ensemble testing of the complete system, covering the roles of prosumers, devices, and the REC aggregator. By enabling seamless communication and coordination between these different actors, the proposed framework will facilitate more efficient participation in flexibility markets and contribute to the stability and quality of the electricity grid.

**Declaration of Competing Interest**

The authors declare that they have no known competing financial interests or personal relationships that could have appeared to influence the work reported in this paper.

**Data availability**

Data will be made available on request.

**Acknowledgements**

This work was supported in part by the PROMINT-CM Project by the Comunidad de Madrid and the European Social Fund under Grant S2018/EMT-4366, in part by the COPILOT-CM Project by the Comunidad de Madrid under Grant Y2020/EMT-6368 and in part by the RuralVPP project ref. SBPLY/21/180501/000223 funded by the Junta de Comunidades de Castilla-La Mancha (JCCM) and the European Union through the European Regional Development Fund.